\documentclass{smjour}
\usepackage{rstruct,psfig}

\newtheorem{fact}{Fact}
\def\beginproof{\noindent{\bf Proof.}\quad}
\def\endproof{}
\def\NP{\rm NP}
\def\AND{\wedge}
\def\OR{\vee}

\def\goesto{\rightarrow}

\def\N{{\bf N}}
\def\PR{{\rm Pr}}
\def\HSAT{{\rm HORN}\hbox{-}{\rm SAT}}
\def\PUR{{\rm PUR}}

\begin{document}




\authorrunninghead{G. Istrate}
\titlerunninghead{Random Horn Satisfiability}

\setcounter{page}{1} 


\title{The phase transition in random Horn satisfiability and its algorithmic implications}

\author{Gabriel Istrate}

\affil{CIC-3 Division and Center for Nonlinear Studies\\
        Los Alamos National Laboratory\\
        Los Alamos, NM 87545, U.S.A.\\
        {\tt gistrate@cnls.lanl.gov}}

\abstract{
Let $c>0$ be a constant, and $\Phi$ be a random Horn 
formula with $n$ variables
and $m=c\cdot 2^{n}$ clauses, chosen uniformly at random (with
repetition) from the set of all nonempty  
Horn clauses in the given variables. 
By analyzing \PUR, a natural implementation of positive unit
resolution, we
show that $\lim_{n\goesto \infty} \PR (\mbox{$\Phi$ is satisfiable})= 1-F(e^{-c})$,
where $F(x)=(1-x)(1-x^2)(1-x^4)(1-x^8)\cdots $. Our method also
yields as a byproduct an average-case analysis of this algorithm. 
}

\section{Introduction}\label{sec:intro}
{\em Phase transitions in combinatorial problems} were first
displayed in the seminal work of Erd\H{o}s and
R\'enyi~\cite{erdos:renyi} on
random graphs.  Working with the constant probability model $G(n,p)$
they showed that the probability that the
graph has a ``large'' connected component exhibits a sharp increase at
some ``threshold''  value of $p$. The empirical
observation from \cite{cheeseman-kanefsky-taylor}, that for a number
of \NP --complete problems 
the ``hardest on the average'' instances are located near such
threshold points has attracted considerable interest in such threshold 
phenomena from several communities, such as Theory of Computing, Artificial
Intelligence and Statistical Mechanics. Recent studies \cite{2+p:rsa,2+p:nature} have provided
further evidence that (at least some) phase transitions have
indeed an impact on algorithmic complexity, and have offered
additional insight on the cases when this happens. 

It turns out that there are two different notions of phase transition 
in a combinatorial problem $P$. One definition applies to optimization 
problems and directly parallels the approach from Statistical Mechanics. 
Potential solutions for an instance of $P$ are viewed as ``states'' of 
a system. One defines an abstract {\em Hamiltonian (energy) function}, 
that measures the ``quality'' of a given solution, and apply methods
from the theory of spin glasses \cite{virasoro-parisi-mezard} to make predictions on the typical
structure of optimal solutions. In this setting a
phase transition is defined as non-analytical behavior of a certain 
``order parameter'' called free energy,
and a discontinuity in this parameter, manifest by the sudden
emergence of a {\em backbone} of constrained ``degrees of freedom''  
\cite{2+p:rsa} is responsible
for the exponential slow-down of many natural algorithms.   

The second definition is combinatorial and pertains to decision
problems. It is the concept of {\em threshold property} from 
random graph theory, more precisely a restricted version of this
notion, called {\em sharp satisfiability threshold}. 
A satisfiability threshold always exists for monotone problems \cite{bollob-thomasson}, but may or may
not be sharp (we speak of a {\em coarse threshold} in the latter
case). It is this notion of phase transition that we are concerned
with in this paper. 
 
From the practical perspective of \cite{cheeseman-kanefsky-taylor} phase transitions are most appealing
in problems that are thought to be ``hard'', in particular, in
\NP --complete problems.  Therefore a lot of recent work  
has been directed towards locating phase transitions in such problems.  
In some cases, the most proeminent of which is Hamiltonian
cycle~\cite{hamcyclerand}), a complete analysis has been obtained. In
other (e.g., 3-SAT~\cite{frieze-suen,kranakis-3sat,
  achlioptas:3sat:pie, janson-et-al:3sat} and
graph-coloring~\cite{chvatal-color,achlioptas-molloy}), obtaining such
an analysis is hard, and indeed not yet accomplished task: for these
problems there exists a fairly large gap between the best rigorous lower
and upper bounds, and the methods that were used to obtain these
bounds do not seem to be capable to yield a tight analysis. 

Understanding the reasons that make problems with similar computational
complexity differ so much with respect to their ``mathematical
tractability'' is clearly a topic worth investigating. A natural 
intuitive explanation of this discrepancy
is that problems that are easy to analyze ``coincide with high
probability'' with problems with a simple ``local''
structure, while problems that are ``hard to analyze'' lack such an
approximation. Such is the case, for instance, of the above mentioned 
Hamiltonian cycle, that ``coincides with high probability'' with the graph
property ``having minimum degree two'' \cite{aks:hamcycle-hitting}. 
Support in favor of this
intuition also comes from Friedgut's result on the existence of a sharp 
threshold for 3-SAT \cite{friedgut:k:sat}: his proof relies on
showing that problems with coarse thresholds can be well approximated
by some simple ``local'' property, and then proving that 3-SAT lacks such
an approximation. While his result sheds no light on the
``mathematical tractability'' of Hamiltonian cycle, it is tempting to
speculate that there might be a suitable generalization of the concept
of ``coarse threshold'', that 3-SAT still lacks, and that encompasses
all known ``mathematically tractable cases''.   

A natural testbed for the above intuition is the case of polynomial 
time solvable problems. In these cases the hypothesis predicts that
one should be able to obtain a complete analysis: often 
tractability arises from the
existence of a ``local'' characterization, that circumvents the need
for exhaustively searching the exponentially large space of potential solutions.  
Another reason is methodological: studying tractable problems
usually amounts to probabilistic analyses of decision algorithms 
for these problems using a methodology based on Markov chains, a task  
that can often be accomplished.     

Such an approach was successful for some tractable versions of propositional
satisfiability: out of the six maximally tractable
cases of SAT that Schaefer identified in his celebrated Dichotomy 
Theorem~\cite{schaefer-dich}, two are trivially satisfiable and
two have completely analyzed phase transitions. The transition for
2-SAT, the satisfiability problem for CNF formulas
with clauses of size two, has been studied
in~\cite{mickgetssome,goerdt:2cnf} and 
that for XOR-SAT, the satisfiability problem for linear systems of
equations with boolean variables, has been studied
in~\cite{xorsat}). The remaining two cases are the Horn formulas and the 
negative Horn formulas (which are, of course, dual).

In this paper we deal with these two cases. 
Unlike the other two nontrivial cases, we show that Horn satisfiability has 
a {\em coarse threshold}. In the ``critical region'' the 
number of clauses is exponential in the number of
variables, hence from a practical perspective, our results show that 
if do not restrict clause length, random Horn formulas of practical
interest are almost certainly satisfiable (we have subsequently
analyzed the bounded clause length case in \cite{istrate:stoc99}). 
Also, we obtain our result by modeling \PUR, 
a natural implementation of {\em positive unit resolution}, 
by a Markov chain, and our method yields as a byproduct 
an average-case analysis of this algorithm. 

\section{Results}
A {\em Horn clause} is a disjunction of literals containing {\em at
most one positive literal}. It will be called {\em positive} if it
contains a positive literal and {\em negative} otherwise. 
A Horn formula is a conjunction of Horn
clauses. {\em Horn satisfiability} (denoted by $\HSAT$) is the
problem of deciding whether a given Horn formula has a satisfying
assignment.

Since our main interest is in phase transitions in decision problems in
the class NP, we will discuss the notion of satisfiability threshold 
in the framework of {\em \NP --decision problems}. Our 
definition is slightly different from the standard one (e.g. \cite{papad:b:complexity}), and
accommodates the fact that legal encodings of instances of a problem
have in general lengths from a restricted set of values.  

\begin{definition}
An {\em \NP --decision problem} is a five-tuple $P=(\Sigma,D,f,g)$ such that 
\begin{enumerate}
\item $\Sigma$ is a finite alphabet. 
\item $f,g:{\bf N}\goesto {\bf N}$ are polynomial time computable,
  polynomially bounded functions. In addition $f$ has range
  $\{0,1\}$. A length $n$ is called {\em admissible} if $f(n)=1$.  
\item $D\subset \Sigma^{*}\times \Sigma^{*}$ is a polynomial time
  computable relation. 
\item for every pair $(x,y)\in \Sigma^{*}\times \Sigma^{*}$, if  $(x,y)\in
  D$ then the length of $x$ is acceptable and $[|y|\leq g(|x|)]$. 
\end{enumerate}

A string $x$ having an admissible length will be called {\em
  an instance of $P$}. A string $y$ such that $(x,y)\in D$ is called
  {\em a
  witness for $x$}, and we write $x\in P$ to state the fact that there
  exists a witness for the instance $x$. Finally problem $P$ is {\em
  monotonically decreasing} if for every instance $x$ of $P$ and every
  witness $y$ for $x$, $y$ is a witness for every instance $z$
  obtained by turning some bits of $x$ from 1 to 0. Monotonically
  increasing problems can be similarly defined. 
\end{definition} 

The three standard probabilistic models from random graph theory \cite{bol:b:random-graphs}, 
the constant probability model, the counting model, the multiset model 
extend directly to any \NP --decision problem, and are equivalent under 
fairly liberal conditions. For the purposes of this paper we recall the 
definition of the multiset model: 

\begin{definition}
Let $P$ be an \NP --decision problem 
The {\em random multiset model} $\overline{\Omega}(n,m)$ has two parameters, an
{\em admissible length $n$} and an {\em instance density} $1\leq m \leq n$. A random sample 
$x$ from $\Omega(n,m)$ is an instance of $P$ obtained by first setting $x=0^{n}$, then choosing, uniformly at random and
with repetition, $m$ bits of $x$ and switching them
to $1$. 
\end{definition}

Next we define out threshold properties for monotonically decreasing
problems under the multiset model. 
Similar definitions can be given for monotonically increasing
problems, or when using one of the two other random models. 

\begin{definition}
Let $P$ be any monotonically decreasing 
decision problem under the multiset random model
$\Omega(n,m)$. A function $\overline{\theta}$ is a
{\em threshold function for $P$} 
if for every function $m$, defined on the set of admissible instances
and taking integer values, we have
\begin{enumerate}
\item
if $m(n)=o(\overline{\theta}(n))$ then
$\lim_{n\goesto \infty}
\PR_{x\in \Omega(n,m)}[x\in P]=1$,
and 
\item
if $m(n)=\omega(\overline{\theta}(n))$ then
$\lim_{n\goesto \infty}
\PR_{x\in \Omega(n,m)}[x\in P]=0$,

$\theta$ is called {\em a sharp threshold} if in addition the following
property holds:

\item For every $\epsilon >0$ define the two functions $\mu_{1}(n),
  \mu_{2}(n)$ by 
\[ \mu_{1}(n)=\min\{m\in {\bf N}: \PR_{x\in \Omega(n,m)}[x\in P]\leq 1-\epsilon\},
\]
 
\[
\mu_{2}(n)=\min\{m\in {\bf N}: \PR_{x\in \Omega(n,m)}[x\in P]\leq \epsilon\}.
\]
 Then we have  
\[
\\lim_{n\goesto \infty}\frac{\mu_{2}(n)-\mu_{1}(n)}{\overline{\theta}(n)}=0.
\] 
\end{enumerate}

If, on the other hand, for some $\epsilon >0$ the amount
$\frac{\mu_{2}(n)-\mu_{1}(n)}{\overline{\theta}(n)}$ is bounded away from 0
as $n\goesto \infty$,  $\overline{\theta}$ is called a {\em coarse threshold}. These
two cases are not exhaustive as the above quantity could in principle
oscillate with $n$. Nevertheless they are so for most ``natural''
problems.  
\end{definition}

A useful modification of the above framework has the set of admissible
lengths specified by an increasing function $N:{\bf N}\goesto {\bf
  N}$. We correspondingly redefine the random model as
$\Omega(n,m)=\overline{\Omega}(N(n),m)$ and the threshold function by 
$\theta(n)=\overline{\theta}(N(n))$. Such will be the case of 
random Horn satisfiability, for which 
a random formula from $\Omega(n,m)$ is
obtained by choosing $m$ clauses independently,
uniformly at random and with repetition from the set of all $N(n) =
(n+2)\cdot 2^{n}-1$ Horn
clauses over variables $x_{1}, \ldots, x_{n}$. 

The following is our main result:

\begin{theorem}\label{maintheorem}
$\theta(n)= 2^{n}$ is a threshold function for random Horn
satisfiability.

Moreover, for every constant $c>0$ 
\begin{equation}\label{formula}
\lim_{n\goesto \infty}
\PR_{\Phi \in \Omega(n,c\cdot2^{n})}[ \Phi \mbox{ is satisfiable} ] = 1-F(e^{-c}),
\end{equation}
where
\[
F(x) = (1-x)(1-x^2)(1-x^4)\cdots
(1-x^{2^{k}})\cdots.
\]
\end{theorem}

The result makes clear that random Horn satisfiability has a {\em coarse threshold}.

\begin{figure}
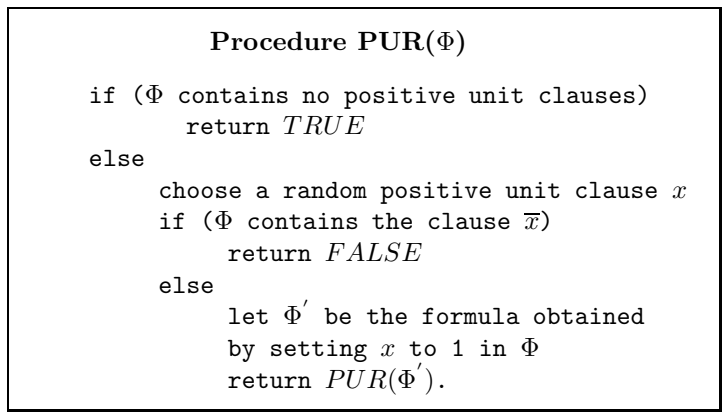
\label{pur}
\begin{algorithm}
\algoindent{25pt}
\begin{centercodebox}{.75 \textwidth}\tt
        {\bit Procedure PUR($\Phi$)}

if ($\Phi$ contains no positive unit clauses)
       return $TRUE$
else 
     choose a random positive unit clause $x$
     if ($\Phi$ contains the clause $\overline{x}$)
          return $FALSE$
     else 
          let $\Phi^{'}$ be the formula obtained 
          by setting $x$ to 1 in $\Phi$
          return $PUR(\Phi^{'})$.

\end{centercodebox}
\end{algorithm}
\caption{Algorithm PUR}
\end{figure}

The algorithm $\PUR$, employed in the proof Theorem~\ref{maintheorem} is
displayed in Fig. 1. $\PUR$ is a natural implementation of positive
unit resolution, which is complete for HORN-SAT \cite{unit-resolution}.

As a byproduct, our analysis yields the following two results, which
provide an average-case analysis of \PUR: 
 
\begin{theorem}\label{sat}
Let $X_{n}\in [0,n]$ be the r.v. denoting the number of iterations of \PUR\ on
a random {\em satisfiable} formula $\Phi\in \Omega(n,c\cdot
2^{n})$. Then $X_{n}$ converges in distribution to a distribution 
$\rho$ on $[0,n]$ having support on the nonnegative integers, $\rho=(\rho_{k})_{k\geq
0}$, $\rho_{k}= Prob[\rho = k]$, 
given by
\[ \rho_{k}=\frac{e^{-2^{k}c}}{1-F(e^{-c})}\cdot \prod_{i=1}^{k-1}
(1-e^{-2^{i}c}).
\]
\end{theorem}

The case of unsatisfiable formulas displays one feature not present
in the previous result: fluctuations due to the
nature of the binary expansion of $n$, {\em wobbles} in the
terminology of P. Flajolet \cite{flajolet:aofa}. 
 
\begin{theorem}\label{unsat}
Let $Y_{n}$ be the r.v. denoting the number of iterations of \PUR\ on
a random formula $\Phi\in \Omega(n,c \cdot 2^{n})$, and, for
 $k\in [0,n]$, possibly a function of $n$, let $\eta_{n,k}$ 
be the probability that $Y_{n}=\lfloor \log_{2} n \rfloor +k$,
conditional on $\Phi$ being unsatisfiable. Then
\begin{itemize}
\item $\lim_{n\goesto \infty}|k-\log_{2}(n)|=\infty$ implies
  that $\lim_{n\goesto \infty}\eta_{n,k}=0$ 
\item for every $k\in {\bf Z}$
\[\eta_{n,k} = G(k-1,c_{n})-G(k,c_{n})+o(1),
\]
where 
\[ G(k,c)= e^{-c (\sum_{j=-\infty}^{k}2^{j})},
\]
\[c_{n}=\frac{c}{2^{\{\log_{2} (\sqrt n)\}}}.
\]
\end{itemize}
\end{theorem}

\section{Notation and useful results}
For $n\in \N$ and $0\leq p\leq
1$, we denote by $B(n,p)$ a random variable
having a Bernoulli distribution with parameters $n,p$. For $\lambda\in
{\bf R}$, $Po(\lambda)$ will denote a Poisson distribution with expected
value $\lambda$.
 
We will use ``with high probability'' (w.h.p.)
as a substitute for ``with probability $1-o(1)$.''
We also say that a sequence $(p_{n})_{n\in \N}$ of real numbers
is {\em exponentially small} (written $o(1/poly)$) if
for every polynomial $Q$, $p_{n}=o(1/Q(n))$.        
We will measure, as usual, the distance between two probability
distributions with integer values $P=(p_{i})$ and $Q=(q_{i})$ by their
{\em total variation distance} $d_{TV}(P,Q)= \frac{1}{2}\cdot
\sum_{i}|p_{i}-q_{i}|$, and recall the following inequalities from
\cite{sheu:poisson} and \cite{barbour:holst:janson} (page
2 and Remark 1.4):
 
\begin{lemma}\label{b:h:j}If $n,p,\lambda, \mu >0$ then
\[d_{TV}(B(n,p),Po(np))\leq \min\left\{np^{2},\frac{3p}{2}\right\}
\]
 
\[d_{TV}(Po(\lambda), Po(\mu))\leq |\mu - \lambda|.
\]
 
\end{lemma}
 
\begin{definition}
Given two probability distributions $D$ and $D^{\prime}$, we say that {\em
$D^{\prime}$ stochastically dominates $D$} if for every  $x$, $\Pr[D\geq x]
\leq \Pr[D^{\prime}\geq x]$, and write $D\prec D^{\prime}$ when this holds.
\end{definition}
                 
The following are two conditional probability tricks.
\begin{fact}\label{trick-approx}
Let $A_{n}, B_{n}$, and $C_{n}$ be events such that
$\PR[ C_{n}|B_{n} ]=1-o(1)$.  Then
$$|\PR[A_{n}|B_{n}]-\PR[A_{n}|B_{n}\AND C_{n}]|=o(1).$$
\end{fact}
 
\beginproof
 
Applying the chain rule for conditional probability we get
\begin{eqnarray*}
|\PR[A_{n}|B_{n}]-\PR[A_{n}|B_{n}\AND C_{n}]| & = & \\
 | \PR[A_{n}|B_{n}\AND
C_{n}]\cdot \Pr[C_{n}|B_{n}]+\PR[A_{n}|B_{n}\AND
\overline{C_{n}}]\cdot \Pr[\overline{C_{n}}|B_{n}]-\PR[A_{n}|B_{n}\AND
C_{n}]|
& = & \\
| \PR[A_{n}|B_{n}\AND
C_{n}]\cdot(1-o(1))+ \PR[A_{n}|B_{n}\AND
\overline{C_{n}}]\cdot o(1)-\PR[A_{n}|B_{n}\AND C_{n}]| = o(1). 
\end{eqnarray*}
 
 \qed
 
\begin{fact}\label{trick-max}
If $B$ is a random variable taking integer values in the interval $I$,
then for every event $A$,
\[
\min_{\lambda \in I}\{ \PR[ A|(B=\lambda) ] \}
\leq
\PR[A]
\leq
\max_{\lambda \in I}\{ \PR[ A|(B=\lambda) ] \}.
\]
\end{fact}
 
Several ``concentration of measure'' results will be used in the
sequel. They include:
 
\begin{proposition}(Chernoff bound)\label{chernoff}    
Let $X_{1}, \ldots , X_{n}$ be independent 0/1 random variables with
$Pr(X_{i}=1)=p$. Let $X=X_{1}+\ldots +X_{n}$, $\mu = E[X]$ and $\delta
>0$. Then
 
$$Pr[|X-\mu|\geq \delta \cdot \mu]\leq \left[\frac{e^{\delta}}{(1+\delta)^{1+\delta}}\right]^{\mu}.$$
 
\end{proposition}                    

A related inequality from \cite{probabilistic-method} is:  
\begin{proposition}\label{chernoff:poisson}
Let $P$ have Poisson distribution with mean $\mu$. For $\epsilon >0$,
 
\[ \Pr[P\leq \mu \cdot (1-\epsilon)] \leq e^{\epsilon^{2}\cdot \mu
/2},
\]
 
\[ \Pr[P\geq \mu \cdot (1+\epsilon)] \leq
[e^{\epsilon}(1+\epsilon)^{-(1+\epsilon)}]^{\mu}.
\]
\end{proposition}             

We regard the algorithm $\PUR$ as working in stages, indexed by the
number of variables still left unassigned; thus the stage number
decreases as $\PUR$ moves on.  Let $\Phi$ denote an input formula over
$n$ variables.
For $i, 1\leq i\leq n$, $A_i$, $R_i$, and $S_i$ respectively denote the
event that $\PUR$ accepts at stage $i$,
the event that $\PUR$ rejects at stage $i$,
and the event that $\PUR$ reaches stage $i-1$ (``survives stage $i$''). 
Also, $\Phi_i$ denotes
the $\Phi$ at the beginning of stage $i$, $N_i$ denotes the number
of clauses of $\Phi_i$, $HP_{1,i}$ the number of positive unit clauses
of $\Phi_{i}$, $HP_{2,i}$ the number of positive {\em non-unit}
clauses, $HN_{1,i}$ the number of negative unit clauses
and $HN_{2,i}$ the number of negative non-unit clauses.
Finally, for simplicity define $\Pi=F(e^{-c})$ and $\Pi_{i}$
to be the product of the first $i$ terms from $\Pi$.    
   
We will assert stochastic domination via {\em couplings of Markov
chains} (for an extensive treatment see \cite{lindvall:coupling}). 
The framework needed for our coupling result is made 
precise in the following definitions (especially tailored for the
context of this paper, rather than being standard).

\begin{definition}
Let $(X_{n})$ be a Markov chain having state space $S$ and transition
matrix $X$.  
A {\em stopping rule $H$ for $X_{n}$} is a set $H$ of {\em transitions
  of $(X_{n})$} (i.e. pairs of states $(i,j)\in S\times S$ such that
$X_{i,j}>0$). 
\end{definition} 

{\underline{\bf Intuition:}}\hspace{5mm}  We will use stopping rules $H$ to talk
about the probability (denoted $\Pr[A|H]$) 
of properties $A$ of the Markov chain that only hold conditional on
$(X_{n})$ making only transitions from $H$. 

\begin{definition}
Let $X_{t}=(X_{0,t},\overline{X}_{t})$ and $Y_{t}=(Y_{0,t},\overline{Y}_{t})$ be two Markov
  chains on ${\bf Z}\times {\bf Z}^{d}$ having transition matrices
  $X$, $Y$, respectively. Let $H_{1}$, $H_{2}$ be two stopping rules
  for $(X_{n})$, $(Y_{n})$, respectively. Let $0\in B\subset\{0,\ldots,
  d\}$.  
A {\em $(B,H_{1},H_{2})$-majorizing (Markovian) coupling of $X$ and $Y$} is a Markov chain $Z=(Z_{t,1},Z_{t,2}))$ on 
  $({\bf Z}\times {\bf Z}^{d})^{2}$, $Z_{t,1}=(Z_{t,01},\ldots, Z_{t,d1})$,
  $Z_{t,2}=(Z_{t,02},\ldots, Z_{t,d2})$, having transition matrix 
$(Z_{(i,j),(k,l)})_{i,j,k,l\in {\bf Z}^{d+1}}$  such that:
\begin{itemize}
\item for every $i,j\in {\bf Z}^{d+1}$, $\Pr[Z_{t+1,1}]=j|Z_{t,1}=i]=X_{i,j}$,
\item for every $i,j\in {\bf Z}^{d+1}$,
  $\Pr[Z_{t+1,2}]=j|Z_{t,2}=i]=Y_{i,j}$,
\item for every $i,j,k,l\in {\bf Z}^{d+1}$, if $Z_{(i,j),(k,l)}>0$ and
  $(i,k)\in H_{1}$ then $(j,l)\in H_{2}$. 
\item for every $t\geq 0$ and every state
$(Z_{t,1},Z_{t,2})$ of $Z_{t}$ reachable through moves in $H_{1}\times
({\bf Z}^{d+1})^{2}$ only, we have  
\[
Z_{t,i1}=Z_{t,i2}\mbox{ for all } i\in B,
\]

and 

\[
Z_{t,01}\leq Z_{t,02}. 
\] 

\end{itemize}
\end{definition}
 
{\underline{\bf Intuition:}}\hspace{5mm} The first two conditions
express the fact that the coupling is {\em Markovian}. 
The third condition (denoted symbolically $H_{1}\leq H_{2}$)
relate the two stopping rules. Finally, the last condition allows us
to compare two quantities of interest for the Markov chains $(X_{n})$
and $(Y_{n})$, namely $\sum_{i\in B}X_{i,t}$ and $\sum_{i\in
  B}Y_{i,t}$. 

Let us now formally state this comparison result. 

\begin{lemma}\label{maj:coupling}
Let $(X_{t})$, $(Y_{t})$, $H_{1}$, $H_{2}$, $B$ be as in the previous
definition, and suppose it is possible to construct a
$(B,H_{1},H_{2})$-majorizing coupling of $(X_{t})$ and
$(Y_{t})$. Then, for every $a\in {\bf Z}$, 

\[
\Pr[\sum_{i\in B}X_{i,t}\geq a | H_{1}]\leq \Pr[\sum_{i\in B}Y_{i,t}\geq a | H_{2}]
\]
\end{lemma}
\beginproof

Define 
\[
H_{B,a}=\{\lambda=(\lambda_{0},\ldots, \lambda_{d}): \sum_{i\in
  B}\lambda_{i}\geq a\}.
\]

Then 

\begin{eqnarray}
\Pr[X_{t}\in H_{B,a}| H_{1}] & = & \sum_{x\in H_{B,a}}\Pr[X_{t}=x|H_{1}]\label{e0}\\ 
& = & \sum_{x\in H_{B,a}}\Pr[Z_{t,1}=x|H_{1}\times S^{2}]\label{e1}\\ 
& = & \sum_{x\in H_{B,a}}\sum_{y\in S} \Pr[(Z_{t,1}=x) \AND
(Z_{t,2}=y)|H_{1}\times S^{2}]\label{e2}\\
& = & \sum_{x\in H_{B,a}}\sum_{y\in S} \Pr[(Z_{t,1}=x) \AND
(Z_{t,2}=y)|H_{1}\times H_{2}]\label{e3}\\
& = & \sum_{x\in H_{B,a}}\sum_{y\in H_{B,a}} \Pr[(Z_{t,1}=x) \AND
(Z_{t,2}=y)|H_{1}\times H_{2}]\label{e4}\\
& = & \sum_{y\in H_{B,a}}\sum_{x\in H_{B,a}} \Pr[(Z_{t,1}=x) \AND
(Z_{t,2}=y)|H_{1}\times H_{2}]\label{e5}\\
& \leq & \sum_{y\in H_{B,a}}\sum_{x\in S} \Pr[(Z_{t,1}=x) \AND
(Z_{t,2}=y)|H_{1}\times H_{2}]\label{e6}\\
& \leq & \sum_{y\in H_{B,a}}\sum_{x\in S} \Pr[(Z_{t,1}=x) \AND
(Z_{t,2}=y)|S^{2}\times H_{2}]\label{e7}\\
& = & \sum_{y\in H_{B,a}}\Pr[Z_{t,2}=y|S^{2}\times H_{2}]\label{e8}\\
& = & \Pr[Y_{t}\in H_{B,a}| H_{2}]\label{e9}.
\end{eqnarray}

Lines~\ref{e1},~\ref{e9} follow from the Markovian character of the
coupling. Line~\ref{e3} follows from $H_{1}\leq H_{2}$. The rest are
simple arithmetical calculations. 
 
\endproof
\qed

The couplings we need are very simple, and employ the following idea: suppose
the recurrences describing $X_{t+1}-X_{t}$ and $Y_{t+1}-Y_{t}$ are identical,
except for one term, which is $B(m_{1},\tau)$ for  $(X_{t})$ and
$B(m_{2},\tau)$ for $(Y_{t})$,
where $m_{1}\leq m_{2}$ are positive integers and $\tau \in (0,1)$.
Obtain a coupling by identifying $B(m_{1},\tau)$ with the outcome of
the first $m_{1}$ Bernoulli experiments in $B(m_{2},\tau)$.
                           
\section{The Uniformity Lemma}
 
The crux of our analysis relies on the observation  
that the behavior of $\PUR$ on a random Horn instance can be
described by a stochastic recurrence (Markov chain).
 
\begin{figure}
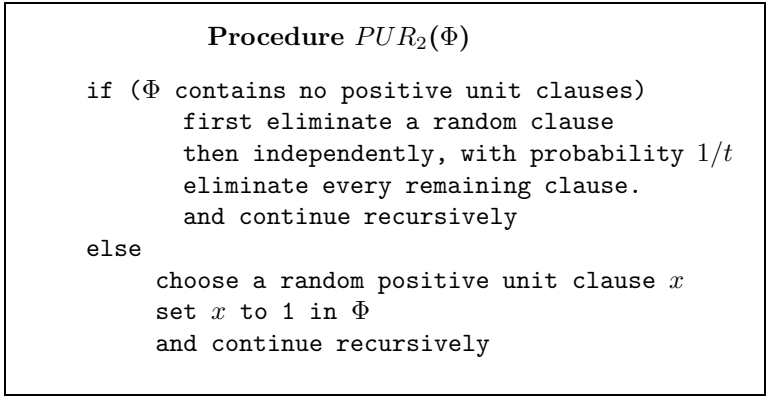
\label{pur:2}

\begin{algorithm}
\algoindent{25pt}
\begin{centercodebox}{.8 \textwidth}\tt
        {\bit Procedure $PUR_{2}$($\Phi$)}

if ($\Phi$ contains no positive unit clauses)
       first eliminate a random clause
       then independently, with probability $1/t$ 
       eliminate every remaining clause. 
       and continue recursively
else 
     choose a random positive unit clause $x$
     set $x$ to 1 in $\Phi$ 
     and continue recursively

\end{centercodebox}
\end{algorithm}
\caption{Second version of PUR}
\end{figure}

\begin{lemma}\label{uniformity}({\bf ``The Uniformity Lemma'' :})
\begin{enumerate}
\item 
Suppose $\PUR$ does not halt before stage $t$. Then, conditional on $N_{t}$,
the clauses of $\Phi_{t}$ are random and independent.

\item Consider $PUR_{2}$, the modified version of the algorithm $\PUR\ $ from
  Figure 2 (that does not check for accepting/rejecting, but may 
  produce empty clauses). 

Let $E_{i}$ represent the number of empty 
  clauses at stage $i$. Then for every
  stage $t$, conditional on $\Gamma_{t}=(HN_{1,t},HN_{2,t},HP_{1,t}, HP_{2,t}, E_{t})$
  the clauses of $\Phi_{t}$ are chosen uniformly at random and are independent.

\item Consider again the original version of \PUR\ .  
Suppose now that we condition $\Gamma_{t}$ and on the fact that $\Phi$
survives Stage $t$ as well. Then  we have 

\begin{equation}\label{eq:markovchain}
N_{t-1}=N_{t}-\Delta_{1,P}(t)-\Delta_{2,P}(t), 
\end{equation}

where 
\begin{itemize}
\item $\Delta_{1,P}(t)$, the number of positive clauses that are
satisfied at stage $t$, has the distribution $1+B\left(HP_{1,t}-1,\frac{1}{t}\right)$. 
\item  
$\Delta_{2,P}(t)$, the number of positive non-unit clauses 
that are satisfied at stage $t$, has the binomial distribution
$B\left(HP_{2,t},\frac{1}{t}\right)$.
\end{itemize}
\end{enumerate}
\end{lemma}

\beginproof
 
The proof is
based on the {\em method of deferred
decisions}~\cite{deferred:decisions}. 
The crux of this method is to consider the
random formula $\Phi$ as being disclosed gradually as the algorithm 
proceeds, rather than as being completely determined at the very
beginning of the algorithm. Following a suggestion of Achlioptas
\cite{achlioptas:3sat:pie} the process can be conveniently imagined as having the
occurrences of each literal in the formula represented by a card that
has the literal as it value. The cards corresponding to each clause
are arranged in separate
piles, and are all initially face down (to reflect the fact that initially 
we don't know anything about the formula). Part of the unveiling
process will consist of {\em dealing} (turning face up) the cards 
from each pile that contain a specific literal. We also assume that
(unless other specified by the unveiling process) the still undealt
parts of each pile is ``hidden'', so that we don't know its
height. 

\begin{enumerate}
\item 
For the first part of the lemma (that conditions only on $N_{t}$) the 
disclosure process consists of first unveiling, at each stage greater
than $t$, the location of a random positive unit clause of $\Phi_{t}$
(guaranteed to exist). We fill it with a random variable among those
left. The process continues by providing 
\begin{enumerate}
\item all the occurrences of this variable.
\item the locations and complete contents of clauses that contain this
  variable in positive form, and 
\item the locations of the clauses that have been completely
  filled. 
\end{enumerate}
We refer to the clauses in the latter two cases as {\em blocked},
since we have complete information about them, and they will no longer
be involved in the unveiling process.

Suppose $\PUR$ arrives at stage $t$ on $\Phi$.  Then in stages
$i=n, n-1, \ldots, t+1$, $\Phi_i$ should have contained a unit clause
consisting of a positive literal but not its complement. This
information does not condition in any way the structure of the
clauses of $\Phi_{t}$, that correspond to the non-blocked piles, counted by
$N_{t}$. In fact that the only information we have at Stage $t$ about
these piles is their number $N_{t}$.

For each such pile all disclosed literals appear only in negative 
form, since otherwise the clause would have been satisfied and
blocked. Hence the {\em residual} (hidden) part still obeys the Horn
restriction. Given the uniformity in the choice of the initial
clauses of $\Phi$, it follows that the clauses of $\Phi_{t}$ are chosen
uniformly at random (and independently) among all nonempty Horn clauses
in the remaining variables.   

\item 

We will prove the result inductively, starting with Stage $n$ (where
it certainly is true) and working downwards. At each stage, the
disclosure process will offer some information on the type of the 
hidden portion of the clause, namely whether it is a positive unit, positive non-unit,
negative or empty. 

\begin{definition}
For notational convenience define
$p_{1}(t)=\frac{1}{t}$, $p_{2}(t)=\frac{1}{2^{t-1}-1}$, 
$p_{3}(t)=\frac{1}{2}$, 
$p_{4}(t)=\frac{t-1}{(2^{t}-t-1)}$.   
\end{definition}

{\bf If $HP_{1,t}>0$,  to carry on the disclosure process:} 

\begin{enumerate}
\item choose a random positive unit clause, fill it with a random
  variable $x$ among those left, and block.  
\item independently with probability $1/t$ fill any of the remaining
  positive unit clauses with $x$ and block.  

\item for any positive non-unit clause: 
\begin{enumerate} 
\item with probability $p_{1}(t)$ fill
  one entry of the clause with $x$, fill the rest of the clause with a 
  random, non-empty, combination of negated remaining literals and block. 
\item if the first case did not happen then,   
      with probability $p_{2}(t)$,  fill one entry with $\overline{x}$
and set the type of the remaining clause to ``positive unit''. 
\item if the first two cases did not happen then,  with probability
  $p_{3}(t)$,  fill one entry with $\overline{x}$ (but do nothing else).   
\item otherwise do nothing.  
\end{enumerate}

\item for any negative unit clause: 

\begin{enumerate}

\item with probability $p_{1}(t)$ fill one entry of the clause with
  $\overline{x}$, set the type of the remaining clause to
  ``empty''. 
\item otherwise do nothing. 
\end{enumerate}

\item for any negative non-unit clause: 

\begin{enumerate}

\item with probability $p_{4}(t)$ fill one entry of the clause with
  $\overline{x}$ and set the type of the remaining clause to
  ``negative unit''. 
\item if the first case did not happen then, with probability $p_{3}(t)$, fill one entry of the clause with
  $\overline{x}$ (but do nothing else). 
\item otherwise do nothing. 
\end{enumerate}
\end{enumerate}
{\bf In the opposite case, $HP_{1,t}=0$, }  
the disclosure process consists of performing the procedure described
in the algorithm, and additionally filling every eliminating clause
with a random Horn clause in the remaining variables that is not a
positive unit clause.

By a tedious but straightforward case analysis it is easy to see that in both cases the uniformity property carries through to
the next stage. The reason is that in all cases the only information
we disclose about each remaining clause is its type, but not its
content. Moreover, we get the following recurrences for the case
$HP_{1,t}>0$ :

\[
\left\{\begin{array}{l}
        HP_{1,t-1}= HP_{1,t}-1-\Delta_{1,P}(t)+\Delta_{12,P}(t),\\
        HP_{2,t-1}= HP_{2,t}-\Delta_{2,P}(t)-\Delta_{12,P}(t),\\
        HN_{1,t-1}=HN_{1,t}-\Delta_{E}(t)+\Delta_{12,N}(t),\\
        HN_{2,t-1}=HN_{2,t}-\Delta_{12,N}(t),\\
        E_{t-1}= E_{t}+\Delta_{E}(t),
\end{array}
\right.         
\]

where 

\[
\left\{\begin{array}{l}
\Delta_{1,P}(t)=B\left(HP_{1,t}-1, p_{1}(t)\right),\\
\Delta_{2,P}(t)=B\left(HN_{2,t},p_{1}(t)\right),\\
\Delta_{12,P}(t)=B\left(HP_{2,t}-\Delta_{2,P}(t),p_{2}(t)\right)\\
\Delta_{E}(t)=B\left( HN_{1,t}, p_{1}(t)\right),\\
\Delta_{12,N}(t)=B\left(HN_{2,t}, p_{4}(t)\right). 
\end{array}
\right.         
\]

\item The conditioning on \PUR\ surviving Stage $t$ implies that 
up to Stage $t-1$ the algorithm \PUR\ and its modified version
$PUR_{2}$ work
  in the same way. With respect to $PUR_{2}$ 
it gives us one additional piece of information with respect to 
merely conditioning on $\Gamma_{t}$:
that $\Delta_{E}(t)=0$.  The desired recurrence follows from the
previous point.

\end{enumerate}
\endproof
\qed

\subsection{ Comments on the Uniformity Lemma} 

A few comments on the contents of the uniformity lemma are in
order.  Although (as shown by Lemma~\ref{uniformity} (i)) it would seem that we
can characterize the state of \PUR\ at Stage $t$ by a single number,
$N_{t}$, this is not so, for two reasons: 
\begin{itemize}
\item first, the above uniformity result is conditional (on \PUR\ surviving
  Stage $t+1$) and does not hold throughout the whole evolution of the
  algorithm. For instance it is {\em not} true at stages before stage
  $t+1$, since unit clauses that are the negation of the variable
  being set cannot appear.  An unconditional uniformity result is provided by
  Lemma~\ref{uniformity} (ii). However, it applies to a modified algorithm,
  which is no longer complete for \HSAT\ , and cannot be used to
  obtain an exact result (rather than just a lower bound on the threshold, 
  as it is done e.g. in \cite{frieze-suen} for $k$-SAT). 
\item second, as shown by Lemma~\ref{uniformity} (iii),  a 
stochastic recurrence for $N_{t-1}$ {\em cannot} be determined by only 
using the value of $N_{t}$; instead we need additional information on
the structure of $\Phi_{t}$ captured by the five-tuple $\Gamma_{t}$.  
\end{itemize}

Fortunately it is possible to circumvent both these problems. 
On one hand it will turn out that all we need for the analysis is 
the conditional uniformity result (i), as long as we can ``control''
the value $N_{t}$. On the other hand, this value can be indirectly
estimated throughout the ``most interesting regime of \PUR\ ``.

\subsection{A coupling result} 

The following result makes a first step towards
estimating $N_{t}$, by showing that we can ``approximate'' this value
by the value of a Markov chain with a simpler structure. The
intuitive idea is simple: by Lemma~\ref{uniformity} (iii)
the ``net decrease'' $N_{t-1}-N_{t}$ is approximately 
$1+B(HP_{1,t}+HP_{2,t}-1,\frac{1}{t})$ which is intuitively less 
than $1+B(N_{t}-1,\frac{1}{t})$.

\begin{figure}
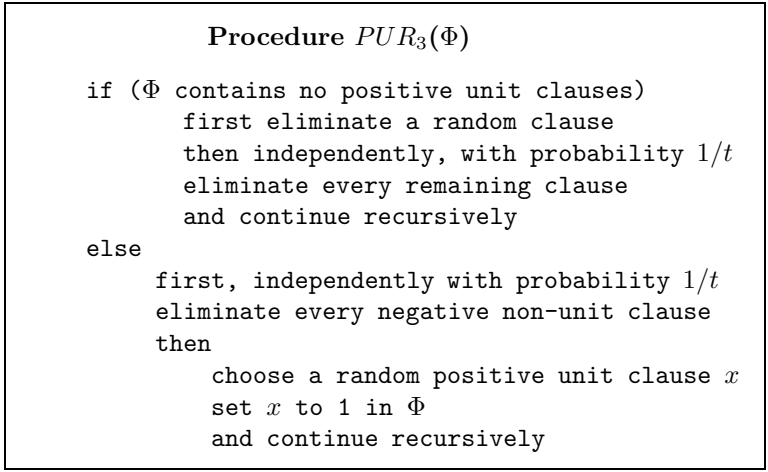
\label{pur:3}
\begin{algorithm}
\algoindent{25pt}
\begin{centercodebox}{0.8 \textwidth}\tt
        {\bit Procedure $PUR_{3}$($\Phi$)}

if ($\Phi$ contains no positive unit clauses)
       first eliminate a random clause
       then independently, with probability $1/t$ 
       eliminate every remaining clause
       and continue recursively         
else 
     first, independently with probability $1/t$
     eliminate every negative non-unit clause
     then 
         choose a random positive unit clause $x$
         set $x$ to 1 in $\Phi$ 
         and continue recursively

\end{centercodebox}
\end{algorithm}
\caption{Third version of PUR}
\end{figure}

\begin{lemma}\label{major}
Consider the modified version of \PUR\ from Figure 3. Then

\begin{enumerate}
\item Conditional on $\Gamma^{(2)}_{t}=(HN^{(2)}_{1,t},HN^{(2)}_{2,t},HP^{(2)}_{1,t}, HP^{(2)}_{2,t}, E^{(2)}_{t})$ (the same quantities as in Lemma~\ref{uniformity}
  (ii); we only use the superscript to indicate the fact that we are
  dealing with a different algorithm) the clauses of $\Phi_{t}$ 
  denote their number by $N^{(2)}_{t}$) are uniform and independent.

\item Define $S_{0}=\{[(a,b,c,d,e)\goesto (a_{1},b_{1},c_{1},d_{1},e_{1})]: 
(c>0)\&\&(e_{1}=0)\}$. Define the stopping rules $H_{2}$, $H_{3}$ for
$\Gamma_{t}$, $\Gamma^{(2)}_{t}$ to be respectively the set of legal transitions
of $\Gamma_{t}$, $\Gamma^{(2)}_{t}$ that are in $S_{0}$. Finally,
define $B=\{0,1,2, 3\}$. 

Then it is possible to construct a $(B,H_{2},H_{3})$--majorizing 
coupling of the Markov chains $\Gamma_{t}$ and $\Gamma^{(2)}_{t}$.

\item  If $HP^{(2)}_{1,t}>0$ then
  $N^{(2)}_{t-1}=N^{(2)}_{t}-1-\Delta_{1,P}(t)-\Delta_{2,P}(t)-
  \Delta_{1,N}(t)-\Delta_{2,N}(t)$, where

\[
\left\{\begin{array}{l}
\Delta_{1,P}(t)=B\left(HP_{1,t}-1, \frac{1}{t}\right),\\
\Delta_{2,P}(t)=B\left(HN_{2,t},\frac{1}{t}\right),\\
\Delta_{1,N}(t)=B\left(HN_{1,t}, \frac{1}{t}\right),\\ 
\Delta_{2,N}(t)=B\left(HN_{2,t}, \frac{1}{t}\right).
\end{array}
\right.         
\]

Consequently, irrespective of the value of $HP^{(2)}_{1,t}$,
\[
N^{(2)}_{t}-N^{(2)}_{t-1}\stackrel{D}{=}1+B(N^{(2)}_{t}-1,
\frac{1}{t}).
\] 

\end{enumerate}
\end{lemma}
\beginproof
\begin{enumerate}
\item 

The proof is identical to the one of Lemma~\ref{uniformity} (ii), and thus
omitted. 

\item 

The intuition behind the definition of the set $S_{0}$ is simple, and
displays the connection with the desired analysis of the algorithm \PUR\ : we
restrict the set of legal transitions of $\Gamma_{t}$,
$\Gamma^{(2)}_{t}$ to those for which $HP_{1,t}>0$ and $E_{t-1}=0$ (in
other words those for which \PUR\ survives stage $t$, and thus works
like $PUR_{2}$). 

The coupling can be described in a very intuitive way. Suppose that
we carry on the disclosure process corresponding to the algorithm
$PUR_{2}$, but the blocking of a clause is accomplished by placing a red
pebble on the corresponding pile, rather than physically eliminating it. We
modify this process to also place, at each stage $j$ such that
$HP_{1,j}>0$, some blue pebbles on the
piles corresponding to negative non-unit clauses, at follows: 
each such clause that has no pebble on it
independently receives a blue pebble with probability $1/j$. It 
is easy to see that the new pebbling process (red and blue) simulates
the algorithm $PUR_{3}$. The coupling easily follows.   

\item 

The result follows from point 1, by separately considering the
behavior of $PUR_{3}$ in the two cases, $HP^{(2)}_{1,t}>0$,
$HP^{(2)}_{1,t}>0$. 

\end{enumerate}
\endproof
\qed

\section{The proof outline}
We will prove only the second part of the theorem, since the first
part directly follows from it.  By the proof of Lemma~\ref{uniformity}
the behavior of the algorithm can be described (with the above
mentioned caveats) by a stochastic
recurrence involving $N_t$. Proposition~\ref{first}
below proves the important fact that with high probability $N_{t}$
stays close to its expected value, which is $N_{n}(1-o(1))$ for
$t=n-O(n^{1/2})$.   

So, intuitively, the number of clauses of $\Phi_{t}$ stays (almost) the
same, while the number of variables decreases by one. The net effect
of one iteration is thus to ``double the constant $c$''.  
We build the proof on three technical
lemmas, Lemmas~\ref{second}, \ref{third}, and \ref{fourth}.
Intuitively, these lemmas show the following:
\begin{itemize}
\item
Lemma~\ref{second} states that with probability $1-o(1)$ $\PUR$
rejects ``in the first $\log n+\theta(1)$ stages'' (if at all;we will make this
more precise in Theorem~\ref{unsat}). 
\item
Lemma~\ref{third} states that with probability $1-o(1)$ $\PUR$ does not reject in any fixed
number of steps.
\item
Lemma~\ref{fourth} obtains a coarse inequality for the satisfaction
probability 
\[    e^{-c}-o(1)\leq \PR[\Phi \in \mbox{HORN-SAT}]
\leq \frac{e^{-c/4}}{1-e^{-c/4}}+o(1).
\]

A consequence of this result
is that a constant number, say $k$, of iterations ``blows up'' $c$
so that the resulting constant $2^{k}c$ is so large that $\Phi_{n-k}$
is unsatisfiable with probability arbitrarily close to 1. 
\end{itemize}

Next we obtain a relation between the probability that $\PUR$ rejects
$\Phi_n$ and the probability that $\PUR$ rejects $\Phi_{n-1}$
($\Phi_{n-1}$ is defined with probability $1-o(1)$ in the case
when $c=\Theta(1)$ due to Lemma~\ref{third}):
the former is equal to the latter multiplied by the probability that
PUR survives stage $n$. This latter term is one minus the probability
that \PUR\ accepts at stage $n$, which is asymptotically equal
$e^{-c}$,  and minus the probability that \PUR\ rejects at step $n$,
which is $o(1)$ and can be asymptotically neglected. 
Iterating this relation for a large enough (but constant) number
of steps $k$ that make $\Pr[\Phi_{n-k}\mbox{ is unsatisfiable}]$
``close enough to 1'' and the partial product $\Pi_{k}$ ``close enough to $\Pi$'' 
allows us to argue that, for every $\epsilon >0$, the probability
that \PUR\ rejects is, for sufficiently large $n$, within $\epsilon$ of
the value $\Pi$ prescribed by the theorem.   
 
\section{The key lemmas}

\begin{proposition}\label{first}
For every $c>0$ and every $t, n-c\sqrt n \leq t \leq n$, 
the conditional probability that the inequality 
\begin{equation}\label{concentrate}
N_{n}-(n-t)\left[1+\frac{2(N_{n}-1)}{t}\right]\leq N_{j}\leq N_{n}\end{equation}
holds for all $t\leq j \leq n$, in the event that $\PUR$ reaches stage $t$,
is $1-o(1)$.
\end{proposition}
 
\beginproof 

For ease of notation, define $E_{t}$ to be the event that Relation~\ref{concentrate} holds, and the sequences
$y_{t}=N_{n}-(n-t)\left[1+\frac{2(N_{n}-1)}{t+1}\right]$ and
$z_{t}=N_{n}$. By the Lemma~\ref{major} (ii) and Lemma~\ref{maj:coupling} we have: 

\[
\Pr[N^{(2)}_{t}\geq y_{t}| H_{3}]\leq \Pr[N_{t} \geq y_{t}|H_{2}]. 
\]

But conditioning on $H_{3}$, $H_{2}$ is the same thing as conditioning
on the algorithms not remaining without unit clauses, and not
producing empty clauses, in other words working like \PUR\ . So 

\[
\Pr[E_{t}| S_{t+1 }]\geq \Pr[N^{(2)}_{t}\geq y_{t}| H_{3}]. 
\]

$H_{3}$ implies that
$N^{(2)}_{j+1}-N^{(2)}_{j}\stackrel{D}{=}B(N^{(2)}_{j+1}-1,\frac{1}{j+1})$
for every $j\geq t$. 

So, defining the Markov chain $U_{t}$ by $U_{n}=N_{n}$ and
$U_{t}-U_{t-1}\stackrel{D}{=}1+\eta_{t}$, where the $\eta_{j}$ are
independent variables having the Bernoulli distribution 
$B(N_{j}-1,\frac{1}{j})$, it follows that 

\begin{equation}\label{et}
\Pr[U_{t}\geq y_{t}]=\Pr[N^{(2)}_{t}\geq y_{t}| H_{3}]\leq \Pr[E_{t}|
S_{t+1}]
\end{equation}

By the Chernoff bound, and reasoning inductively, we infer that with
probability $1-o(1)$ we have $\eta_{j}\leq \frac{2(U_{j}-1)}{j}\leq \frac{2(N_{n}-1)}{t}$ for every 
$t\leq j \leq n$. Plugging this inequality in the
definition of $U_{t}$ and using equation~\ref{et} proves the lemma. 
\endproof
\qed

 
\begin{lemma}\label{second}
Let $p=p(n)$ such that 
$\lim_{n\goesto \infty} [n-\log_{2} n -p(n)]=\infty$.
Then $\PR[R_p|S_{p+1}]$, i.e.,
the conditional probability that $\PUR$ rejects at stage $p(n)$
in the event that $\PUR$ reaches stage $p(n)$, is $1-o(1)$.
\end{lemma}
 
\vspace{5mm}

To prove this lemma we need the following trivial combinatorial
result: 

\begin{lemma}\label{ballsandbins}
 
Let $a(n)$ white balls and $b(n)$ black balls be thrown independently
into $n$ bins. Pick a random bin among those containing a white ball,
and let $X_{n}$ be the event that the chosen bin contains a black ball as well.Then $Pr[X_{n}]=1-(1-\frac{1}{n})^{b(n)}$. 
\end{lemma}

\beginproof

It is easy to see that the bin we choose can be seen as the result of 
choosing a random bin among {\em all} $n$ bins. 
So $\Pr[X]$ is simply the probability
that a randomly chosen bin gets a black ball. But this is 
$1-(1-\frac{1}{n})^{b\cdot n}$. 

\endproof 
\qed

{\bf Proof of Lemma~\ref{second}:}
 
Let $T$ denote the event $E_{n}\AND E_{n-1}\AND \cdots \AND E_{p}$.
It follows from Proposition~\ref{first} that $\PR[T|S_j]=1-o(1)$.
Then, by Fact~\ref{trick-approx},
$\PR[R_p|S_{p+1}]=\PR[R_p|S_{p+1}\AND T]+o(1)$. 
Since $T$ implies $N_{p}\in I= [y_{p}, z_{p}]$, 
\[
\PR[R_p|S_{p+1}\AND T]\geq
\min_{\lambda \in I}\{ \PR[R_p|S_{p+1}\AND T\AND (N_{p}= \lambda)] \}.
\]
Thus, the claim holds if we show that
$\max_{\lambda \in I} \PR[\overline{R_p}|S_{p+1}\AND T\AND (N_{p}= \lambda)] = o(1)$.
 
Suppose that $N_{p}=\lambda$, the events $T$, $S_{p+1}$ hold, 
and we further condition on the number of negative unit clauses.   
The event $R_{p}$ can be mapped into $X_{p}$ of the previous ``balls
into bins''
experiment, with the positive unit clauses representing the white
balls, the negative unit clauses being the black balls, and the
remaining $p$ variables being the bins. 

From Lemma~\ref{uniformity} it follows that the number of 
negative unit clauses of $\Phi_{p}$ has a binomial distribution
$B(\lambda,\frac{p}{N(p)})$. Since $\lambda\frac{p}{N(p)}\geq
y_{p}\frac{p}{N(p)}= (1+o(1))c\cdot 2^{\log_{2}(n)+p(n)}= \omega(n)$,
 it follows
easily by the Chernoff bound that with probability $1-o(1)$ the number
of both positive and negative unit clauses of $\Phi_{p}$ is larger
than $\frac{py_{p}}{2N_{p}}$. Since this amount is $\omega(n)$ the
claim is a consequence of Lemma~\ref{ballsandbins}. 
\endproof
\qed

\begin{proposition}\label{reject1}
With probability $1-o(1)$
$\PUR$ does not reject $\Phi$ 
at stage $n$.
\end{proposition}
 
\beginproof
 
Let $U$ be the number of unit clauses in $\Phi$.
The variable $U$ has a binomial distribution with parameters
$2^{n}c$ and $\frac{2n}{(n+2)2^{n}-1}$, so it is
asymptotically a Poisson distribution with parameter $2c$. 
In fact Proposition~\ref{b:h:j} and  
Proposition~\ref{chernoff:poisson} together imply that with probability $1-o(1)$, 
$U\leq 2c(1+n^{1/3})\leq 4cn^{1/3}$.
 
Consider the $U$ unit clauses of $\Phi$ as being balls to be tossed
into $n$ bins. The probability that two of them end up in the same bin
is at most ${{U}\choose{2}}\cdot \frac{1}{n}$, which, in view of the 
above upper bound on $U$, is $o(1)$. 
So with probability $1-o(1)$ no variable appears more than once in
a unit clause of $\Phi$, and thus, $\PUR$ does not reject.
\endproof

\begin{lemma}\label{third}
For every $k>0$, with probability
$1-o(1)$, $\PUR$ does {\em not} reject in any of the stages
$n, n-1,\ldots, n-k+1$.
\end{lemma}
 
\beginproof
 
A simple induction on $k$, coupled with the fact that, conditioned on
$N_{t}$, $\Phi_{t}$ is a random formula, and Proposition~\ref{concentrate}. 
\endproof
\qed

\begin{lemma}\label{fourth}
For every positive constant $c$,
$e^{-c}-o(1)\leq
\PR[\Phi \in \HSAT]
\leq \frac{e^{-c/4}}{1-e^{-c/4}}+o(1)$.
\end{lemma}
 
\beginproof

Let $c>0$ be a constant.
\begin{eqnarray*}
\PR[\Phi \in \HSAT]
&\geq&
        \PR[\mbox{$\PUR$ accepts at the first step}]    \\
&=&
        \PR[\mbox{$\Phi$ contains no positive unit clauses}]    \\
&=&
        \left(1-\frac{n}{(n+2)\cdot 2^{n}-1}\right)^{2^{n}c}    \\
&=&
        e^{-\frac{n2^{n}\cdot c}{(n+2)\cdot 2^{n}-1}}-o(1)      \\
&\geq&
        e^{-c}-o(1),
\end{eqnarray*}
since $\frac{n2^{n}}{(n+2)\cdot 2^{n}-1}\leq 1$.
This proves the lower bound.
 
In order to prove the upper bound, define
$p=\log_{2} n +\log \log n$,
let $Y$ be the event ``$\PUR$ accepts,'' and
let $Z$ the event
``$\PUR$ stops in at most $p$ iterations.'' By Lemma~\ref{second},
$\PR[Z]=1-o(1)$, so $\PR[Y]\leq \PR[Y|Z]=o(1)$. However, given $Z$, $Y$ is
equivalent to $A_{n}\OR (A_{n-1}\AND S_{n})\OR (A_{n-p+1}\AND
S_{n}\AND \cdots S_{n-p+2})$. So, by the Bayes rule,
$\PR[Y|Z]$ is at most
\[
        \PR[A_{n}]+\PR[A_{n-1}|S_{n}]+\cdots +\PR[A_{n-p+1}|S_{n}\AND
S_{n-1}
\AND
        \cdots \AND S_{n-p+2}].
\]
We cannot apply directly Fact~\ref{trick-approx},
because this sum has an unbounded number of terms.
Instead, we will use the following simple 
consequence of Bayes conditioning:
\[
\PR[A_{i}|S_{n}\AND\cdots \AND S_{i+1}]\leq
\PR[A_{i}|S_{n}\AND\cdots S_{i+1}\AND E_{i}]
 +\PR[\overline{E_i}|S_{n}\AND\cdots S_{i+1}].
\]
From Proposition~\ref{first} the sum of all ``second terms'' is $o(1)$.
As to the first term, the conditioning implies that the clauses of 
$\Phi_i$ are chosen uniformly at random and their number is between
$y_i$ and $z_i$.  Since $\PUR$ accepts $\Phi_i$ if and only if
$\Phi_i$ contains no positive literals, we have
\begin{eqnarray} 
1-\left(1-\frac{i}{(i+2)2^{i}-1}\right)^{y_i}-o(1)
&\leq&
\PR[\overline{A_i}|\overline{S_n}\AND \cdots \AND
        \overline{S_{i+1}}\AND E_{i}] \label{time:acc}            \\
&\leq&
1-\left(1-\frac{i}{(i+2)2^{i}-1}\right)^{z_{i}}+o(1)\nonumber. 
\end{eqnarray}

in particular
 
\[
\PR[A_{i}|S_{n}\AND \cdots \AND S_{i+1}\AND E_{i}]\leq
\left(1-\frac{i}{(i+2)2^{i}-1}\right)^{y_{i}}.
\]
The right hand side is less or equal than
$e^{-\frac{iy_{i}}{(i+2)2^{i}-1}}$. 
Since $\frac{i}{i+2}\geq \frac{1}{3}$ and $y_{i}\geq N_{n}\cdot (1-\frac{\log n +
    \log \log n}{n-\log n +\log \log n}) \geq \frac{3N(n)}{4}$ 
for a sufficiently large $n$
we have, (assuming such an $n$) 
$e^{-\frac{iy_{i}}{(i+2)2^{i}-1}}\leq e^{-\frac{iy_{i}}{(i+2)2^{i}}}\leq 
 e^{-\frac{2^{n-i}c}{4}}$. 

Summing up all these upper bounds for $\PR[A_{i}|S_{n}\AND \cdots \AND
S_{i+1}\AND E_{i}]$ and observing the exponents as part of the
progression $\{\frac{c}{4} \cdot j\}$, we obtain the desired
upper bound $\frac{e^{-c/4}}{1-e^{-c/4}}+o(1)$.
\endproof 
\qed

\section{Putting it all together}
\label{putting:together}
Now we complete the proof of Theorem~\ref{maintheorem} by proving
equation (\ref{formula}).  

In order to prove this result
it suffices to show that 
\begin{equation}\label{limit}
\lim_{n\goesto \infty} \PR_{\Phi \in \Omega(m,n)}
[\mbox{$\PUR$ rejects $\Phi$}]
= F(e^{-c}).
\end{equation}
It is easy to see that $F$ is well-defined on $(0,1)$ and has the
following Taylor series expansion
\[
\tilde{F}(x)=(-1)^{b_{0}}+(-1)^{b_{1}}x+ (-1)^{b_{2}}x^{(2)}+\cdots
(-1)^{b_{i}}x^{i}+ \cdots
\]
with $b_{i}$ being the number of ones in the binary
representation of $i$. Also $F$ is monotonically decreasing, positive on
$(0,1)$, and has limit 1 at 0.

Fix $\epsilon >0$.  Let $R$ be the event ``$\PUR$ rejects
$\Phi$''. 
What we need to show is that
for a sufficiently large $n$,
\begin{equation}\label{epsilonfinal}
(1-\epsilon)\Pi  \leq \PR[R]\leq (1+\epsilon)\Pi.
\end{equation}
Since $\Pi$ converges and $\Pi>0$, there exists some $k_{0}$ such
that for all $k\geq k_{0}$,
\begin{equation}\label{pi:k}
\sqrt{1-\epsilon}<\frac{\Pi_{k}}{\Pi}<(1+\epsilon).
\end{equation}
By Lemma \ref{fourth}, there exist some $n_{0}>0$ and $c_{0}>0$
such that for every $n>n_{0}$ and every $c>c_{0}$,
$\PR_{\Phi\in \Omega(n,2^{n}c)}[\mbox{$\PUR$ rejects $\Phi$}]
>\sqrt{1-\epsilon}$.  Keeping in mind the fact that events 
$A_{n}, A_{n-1}, \cdots, A_{n-k+1}$ are incompatible with $R$
we obtain the equality
\[
\PR[R]
=
\PR[R|\overline{A_n}\AND \cdots \AND \overline{A_{k}}]
\cdot \PR[\overline{A_n}]\cdot
\prod_{1\leq i\leq k}
\PR[\overline{A_{n-i}}|\overline{A_n}\AND \cdots \AND \overline{A_{n-i+1}}].
\]
for every fixed $k$.

Although conceptually simple, the rest of the proof is a little bit
cumbersome. 

We first consider the case $c>4\ln 2$ (so that the upper bound in
Lemma~\ref{fourth} is strictly less than one).

Choose $k$ so that, for large enough $n$, $y_{n-k}>c_{0}\cdot 2^{n-k}$. This is possible 
since $y_{n-k}\geq c\cdot 2^{n}[1-\frac{k}{n-k}]$. 

We claim (and it is in the proof of these two relations where the
assumption $c>4\ln 2$ will be used) that 
for every $j$, $n-k \leq j\leq n$, that
\begin{equation}\label{final-one}
\PR[\overline{A_j}|\overline{A_n}\AND \cdots \AND \overline{A_{j+1}}]
=
\PR[\overline{A_j}|S_{n}\AND \cdots \AND S_{j+1}]+o(1), 
\end{equation}
and
\begin{equation}\label{final-two}
\PR[R|\overline{A_n}\AND \cdots \AND \overline{A_{j+1}}]
=
\PR[R|S_{n}\AND \cdots \AND S_{j+1}]+o(1).
\end{equation}

We will postpone proving these equations and will see how the
theorem can be proven from these equations.
 
From equations~\ref{time:acc} and ~\ref{final-one} it follows that  
\begin{eqnarray} 
1-\left(1-\frac{n-i}{(n-i+2)2^{n-i}-1}\right)^{y_{n-i}} & - & o(1) \\
&\leq &
\PR[\overline{A_{n-i}}|\overline{A_n}\AND \cdots \AND
        \overline{A_{n-i+1}}]             \\
&\leq&
1-\left(1-\frac{n-i}{(n-i+2)2^{n-i}-1}\right)^{z_{n-i}}+o(1)\nonumber. 
\end{eqnarray}

This proves that, for every $i=1, \ldots, k$, 
\[ \lim_{n\goesto \infty} \PR[\overline{A_{n-i}}|\overline{A_n}\AND
\cdots \AND \overline{A_{n-i+1}}] = (1-e^{-c\cdot 2^{i}}).
\] 

In a similar vein, we have, for large enough $n$,
\[
\sqrt{1-\epsilon}
\leq
\PR[R|\overline{A_n}\AND \cdots \AND \overline{A_{n-k+1}}]
\leq 1.
\]

If we take a large enough $n$, since the second part is asymptotically
equal to $\Pi_{k}$, by (\ref{pi:k}) we have (\ref{epsilonfinal}).
 
 For a general $c>0$, define $c^{*}$ to be the infimum of all
$c$'s for which the relation~\ref{limit} holds for every $c^{\prime}>c$. 
Suppose $c^{*}>0$.
The single-step version of (\ref{final-two}) provides
$\PR[R|\overline{A_n}] = \PR[R|S_{n}]+o(1)$, so
$\PR[R]=\Pr[\overline{A_n}]\Pr[R|\overline{A_n}]+o(1)$.
Let $c<c^{*}$ and let $n_{1}$ be such that for all $n\geq n_{1}$,
$2c(1-\frac{1}{n})^{2}>c^{*}$.
By Fact~\ref{trick-approx} and Proposition~\ref{first} we
have $\PR[R|S_{n}] =\PR[R|S_{n}\AND E_{n-1}]+o(1)$.
Then by Fact~\ref{trick-max} we have
$\min_{\lambda \in I}\{\PR[R|S_{n}\AND E_{n-1}\AND (N_{n-1}=\lambda)]\}
        \leq \PR[R|S_{n}\AND E_{n-1}]
        \leq \max_{\lambda \in I}\{\PR[R|S_{n}\AND E_{n-1}\AND (N_{n-1}=\lambda)]\}$. 
Conditioned on surviving stage $n$ and on the value of $N_{i}$,
$\Phi_{n-1}$ is a random formula. Since both $y_{n-1}$ and $z_{n-1}$
are asymptotically equal to $2^{n}c$, for large $n$, $\Phi_{n-1}$ is
a random formula with $n-1$ variables and $2^{n-1}\cdot
(2c+o(1))$ clauses.  Thus, $\lim_{n\goesto
  \infty}\PR[R|S_{n}]=\lim_{n\goesto \infty}\PR[R|S_{n}\AND E_{n-1}]= F(e^{-2c})$.
Since $\PR[\overline{A_n}]$ is asymptotically equal to $1-e^{-c}$, and
$F(c)=(1-e^{-c})F(2c)$, (\ref{epsilonfinal}) holds for $c$. This shows
that $c^{*}=0$, hence ~\ref{limit} is true for every $c>0$.

Now what remains is to prove (\ref{final-one}) and (\ref{final-two}).
We will prove only (\ref{final-one}); proving the other is quite
similar.
Let $T$ be the event that $\PUR$ rejects in one of the first $k$
stages.  Note that $\PR[T]=o(1)$, as seen in Lemma~\ref{third}.
Note that $T=R_{n}\OR (S_{n}\AND R_{n-1})\OR \cdots \OR (S_{n}\AND
\cdots \AND S_{n-k+2}\AND R_{n-k+1})$, so the probability of each of
the $k$ terms in the disjunction is $o(1)$. 
 
Note that
\[
\PR[\overline{A_n}\AND \cdots \AND \overline{A}_{j}]
=\sum_{\stackrel{r\geq j+1}{\epsilon_{r}\in \{-1,+1\}}}\,
\PR[\overline{A_n}\AND \cdots \AND \overline{A_j}\AND
R_{n}^{\epsilon_{n}}\AND \cdots \AND
R_{j+1}^{\epsilon_{j+1}}],
\]
where $X^{-1}$ denotes the opposite of the event $X$. All terms
in the sum, other than $\PR[\overline{A_n}\AND \cdots \AND
\overline{A_j}\AND R_{n}^{-1}\AND \cdots \AND R_{j+1}^{-1}]$
are either inconsistent (the algorithm rejects twice)
or imply one of the terms appearing in the disjunction of
the decomposition of $T$. Thus,
\[
\PR[\overline{A_n}\AND \cdots \AND \overline{A_j}]
=\PR[\overline{A_n}\AND\overline{R_n}\AND \cdots \AND
\overline{A_{j+1}}\AND\overline{R_{j+1}}\AND \overline{A_j}]+o(1),
\]
that is,
\[
        \PR[\overline{A_n}\AND \cdots \AND \overline{A_j}]
        =
        \PR[S_{n}\AND \cdots \AND S_{j+1}\AND \overline{A_j}]+o(1). 
\]
Similarly, $\PR[\overline{A_{n}}\AND \cdots \AND \overline{A_{j+1}}]
        = \PR[S_{n}\AND \cdots \AND S_{j+1}]+o(1)$.

Note that for every sequence of events $A_{n}$ and  $B_{n}$ with
$\liminf_{n\goesto \infty}\PR[B_{n}]>0$,
$|\frac{\PR[A_{n}]+o(1)}{\PR[B_{n}]+o(1)} -\frac{\PR[A_{n}]}{\PR[B_{n}]}|=o(1).$
So, it suffices to show that $\liminf_{n\goesto \infty}\PR[S_{n}\AND \cdots \AND S_{n-k}]>0$.
This probability is
$1- \PR[\PUR$ accepts in one of the first $k$ steps$]$
$ - \PR[\PUR$ rejects in one of the first $k$ steps$]$,
and thus, is at least $1- \frac{e^{-c/4}}{1-e^{-c/4}}-o(1)- \PR[T]$.
Since $\frac{e^{-c/4}}{1-e^{-c/4}}<1$, the required condition is
guaranteed. 

\endproof
\qed 
 
\section{Proof of Theorem 2.2}

From equations (\ref{final-one}) and (\ref{time:acc}) and Proposition~\ref{first} it
follows that the probability that the algorithm accepts {\em exactly
at Stage $k$}, given that it has not stopped before, tends (as
$n\goesto \infty$) to $e^{-2^{k}c}$. We have 

\begin{eqnarray*}
\Pr[A_{n-k}\AND [\Phi \in \mbox{SAT }]] & = &\Pr[A_{n-k}\AND [\Phi \in
\mbox{SAT }]\AND S_{n-k+1}]\\ & = & \Pr[A_{n-k}\AND [\Phi \in
\mbox{SAT }]| S_{n-k+1}]\cdot \Pr[S_{n-k+1}]\\ 
& = &\Pr[A_{n-k}| S_{n-k+1}]\cdot \Pr[S_{n-k+1}].  
\end{eqnarray*}

Therefore 

\begin{eqnarray*}
\rho_{k}=\lim_{n\goesto \infty}\Pr[A_{n-k}|\Phi \in \mbox{SAT }] & = &
 \lim_{n\goesto \infty}\frac{\Pr[A_{n-k}| S_{n-k+1}]}{\Pr[\Phi \in
 \mbox{SAT }]}\cdot \Pr[S_{n-k+1}]\\ & = &\frac{e^{-2^{k}c}}{1-F(e^{-c})}\cdot \prod_{i=1}^{k-1}
(1-e^{-2^{i}c}). 
\end{eqnarray*}

\endproof
\qed

\section{Proof of Theorem 2.3}

We will only provide an outline of the proof of Theorem~\ref{unsat}, sine
its overall philosophy is quite similar to the one used to prove
Theorem~\ref{maintheorem}. 

Redefine, for the purpose of this section, the index $k$ to refer to
events taking place at stage $n-\lfloor \log_{2}(n) \rfloor -k$. For
instance $S_{k}$ is the same as the event $Y_{n}>n-\lfloor \log_{2}(n)
\rfloor -k$. 

Theorem~\ref{unsat} follows, of course, from the following claim
\begin{lemma}\label{approx}
\begin{equation}\label{ultima}
\lim_{n\goesto \infty}\Pr[Y_{n}> \lfloor \log_{2}(n) \rfloor +k|R]-G(k,c_{n})=0.
\end{equation}
\end{lemma}
To prove Lemma~\ref{approx} we first show, using methods similar to
the ones used to prove Lemma~\ref{fourth}, the following result
\begin{lemma}\label{approx2}
\[
\lim_{k\goesto -\infty}\liminf_{n\goesto \infty} \Pr[Y_{n}> \lfloor \log_{2}(n) \rfloor +k|R]=1.
\]
\end{lemma}

The proof of Lemma~\ref{approx} proceeds now by observing that 
\begin{eqnarray*}
\Pr[(Y_{n} & > & \lfloor \log_{2} (n) \rfloor +k) \AND R]\\ 
& = &
\Pr[S_{k}\AND R]  = \Pr[S_{k-1}\AND \overline{R}_{k}\AND R]\\
& = & \Pr[\overline{R}_{k}\AND R| S_{k-1}]\Pr[S_{k-1}]\\
& = & (\Pr[\overline{R}_{k}| S_{k-1}]-o(1))\cdot (\Pr[S_{k-1}\AND
R]+o(1))\\
& = & (\Pr[\overline{R}_{k}| S_{k-1}\AND E_{k}]-o(1))\cdot (\Pr[S_{k-1}\AND
R]+o(1))\\
& = & \Pr[\overline{R}_{k}| S_{k-1}\AND E_{k}]\cdot \Pr[(Y_{n}> \lfloor \log_{2} (n) \rfloor +k-1) \AND R]+o(1)\\
\end{eqnarray*}

By Lemma~\ref{ballsandbins} the first term is approximately $e^{-c_{n}\cdot
  2^{k}}$. 

Iterating downwards for a constant number of steps, up to $k_{0}\in
{\bf Z}$, we infer

\[
\Pr[Y_{n}> \lfloor \log_{2} (n) \rfloor +k| R]= \Pr[Y_{n}> \lfloor
\log_{2} (n) \rfloor +k_{0}|R ]\cdot
\prod_{j=k_{0}+1}^{k}\Pr[\overline{R}_{k}|S_{k-1}\AND E_{k}]+o(1).
\]

Choosing $k_{0}$ small enough so that, by Lemma~\ref{approx2}, the
first term is ``close enough to 1'' and the product is ``close enough
to $G(c_{n},k)$'' proves relation~\ref{ultima}. 

\qed 
\section{Further discussions and open problems}

There are several versions of Horn satisfiability whose phase
transition is worth studying. One of them is the class of 
{\em extended Horn formulas}
\cite{extended:horn:jacm,extended:horn:ipl}, 
for which $\PUR$ is still a valid
algorithm \cite{extended:horn:jacm}. On the other hand, Horn-like
restrictions have been employed to design tractable restrictions of 
various formalisms of interest in Artificial Intelligence, for example
in constraint programming, temporal reasoning, spatial reasoning,
etc. In many such cases positive unit resolution 
has natural analogs, (for instance {\em arc-consistency} in the case 
of ORD-HORN formulas in temporal reasoning \cite{ord-horn}), and
it would be interesting to see whether the ideas in this paper can
inspire similar results. 

Let us also remark that, as shown in \cite{istrate:stoc99}, the
average-case behavior of \PUR\, as displayed in Theorem~\ref{sat},  is
responsible for a physical property called {\em critical behavior}, 
widely studied in Statistical Mechanics and related areas (see, for
instance, \cite{slade:critical}, for the case of percolation), and  
similar to the one observed experimentally in \cite{kirkpatrick:selman:scaling}
for the case of $k$-SAT. 

One final issue is whether one can meaningfully define and study the 
existence of a ``physical phase transition'' in HORN-SAT. The major
problem  is a ``degeneracy'' property of our random model for Horn 
satisfiability: one can satisfy all but the positive unit clauses of
any formula by the assignment $11\ldots 1$. But under
the random model employed in this paper the fraction of such clauses 
is $o(1)$, a property that is not shared by any of the previously
studied problems, and which makes the ``physical interpretation''
problematic. Whether the problem becomes meaningful under a
different random model remains to be seen.

\section*{Acknowledgment}
This paper is part of the author's Ph.D. thesis at the University of 
Rochester. Preliminary conference versions have appeared as
\cite{istrate:aim98} and \cite{istrate:soda99}.  
Support for this work has come from the NSF CAREER Award CCR-9701911 
and the DARPA/NSF Grant 9725021. 

I thank Mitsu Ogihara for substantive conversations that led to
discovering the results in this
paper. I also thank Jin-Yi Cai and Nadia Creignou for 
insightful comments and enjoyable discussions.

\bibliographystyle{alpha}
\bblname{revised}
\articlebibliography{/nh/nest/u/gistrate/bib/bibtheory}

\end{document}